# NaviCell: a web-based environment for navigation, curation and maintenance of large molecular interaction maps

Kuperstein I[1,2,3#], Cohen DPA[1,2,3#], Pook S[4#], Calzone L[1,2,3], Barillot E[1,2,3] and Zinovyev A[1,2,3]*

[1] Institut Curie, 26 rue d'Ulm, 75248 Paris cedex 5, France

[2] INSERM, U900

[3] Mines ParisTech

[4] Sysra

*To whom correspondence should be addressed

#These authors contributed equally to the work

Emails: Inna.Kuperstein@curie.fr, David.Cohen@curie.fr, Stuart.Pook@curie.fr, Laurence.Calzone@curie.fr, Emmanuel.Barillot@curie.fr, Andrei.Zinovyev@curie.fr

## Abstract

**BACKGROUND:** Molecular biology knowledge can be formalized and systematically represented in a computer-readable form as a comprehensive map of molecular interactions. There exist an increasing number of maps of molecular interactions containing detailed and step-wise description of various cell mechanisms. It is difficult to explore these large maps, to comment their content and to maintain them. Though there exist several tools addressing these problems individually, the scientific community still lacks an environment that combines these three capabilities together.

**RESULTS:** NaviCell is a web-based environment for exploiting large maps of molecular interactions, created in CellDesigner, allowing their easy exploration, curation and maintenance. It is characterized by the unique combination of three essential features: (1) efficient map browsing based on Google Maps engine; (2) semantic zooming for viewing different levels of details or of abstraction of the map and (3) integrated web-based blog for collecting the community feedback. NaviCell can be easily used by experts in the field of molecular biology for studying molecular entities of their interest in the context of signaling pathways and cross-talks between pathways within a global signaling network. NaviCell allows both exploration of detailed molecular mechanisms represented on the map and a more abstract view of the map up to a top-level modular representation. NaviCell greatly facilitates curation, maintenance and updating the comprehensive maps of molecular interactions in an interactive and user-friendly fashion due to an imbedded blogging system.

**CONCLUSIONS:** NaviCell provides an easy exploration of large-scale maps of molecular interactions, thanks to the Google Maps and WordPress interfaces, with which many users are already familiar. Semantic zooming which is used for navigating geographical maps is also adopted for molecular maps in NaviCell, making any level of visualization meaningful to the user. In addition, NaviCell provides the framework for community-based curation of maps.

# Background

One of the most important objectives of systems biology is developing a common language for a formal representation of the rapidly growing molecular biology knowledge [1, 2]. Currently, most of the information about molecular mechanisms is dispersed in thousands of scientific publications. This knowledge exists in a human-readable form, which limits its formal analysis by bioinformatics and systems biology tools.

One of the approaches to formalize biological knowledge is to collect the information on molecular interactions in the form of pathway databases [3]. Examples of them are Reactome [4], KEGG [5], Panther [6], SPIKE [7], WikiPathways [8], TransPath [9] and others that are created using various frameworks and formalisms [10]. Most of the pathway databases provide ways for exploring the molecular pathways visually, some include analytical tools for analyzing their structure and some have a possibility to collect users' feedback (see Table 1).

A parallel approach for formalizing the biological knowledge consists in creating graphical representations of the biochemical mechanisms in the form of maps of molecular interactions [11-19]. The idea of mapping some aspects of molecular processes onto a two-dimensional image appeared at the dawn of molecular biology. The first large maps of metabolism, cell cycle, DNA repair have been created manually starting from the '60s and were not supported by any database structure [20, 21]. Such maps can be considered as a collection of biological diagrams, each depicting a particular cellular mechanism, assembled into the seamless whole, where the molecular players and their groups occupy particular "territories". Molecules put close on the map are assumed to have similar functional properties (though it is not always possible to achieve in practice). This geographical metaphor has certain advantages over the database representations for which no global visual image of pathways' functional proximity and crosstalk exist.

A significant achievement of systems biology was in combining both approaches for knowledge formalization into one. For this purpose, it was necessary to develop a graphical language (meaningful to humans), a computer-readable language (onto which the graphical language would be mapped) and a piece of software that would allow charting a large map of molecular interactions and, simultaneously, creating a computer-readable file (serving as a database). This inspired the creation of tools like CellDesigner [22], SBGN standard for graphical representation of biological diagrams [23] and SBML [24] and BioPAX standards for exchanging the content of biological pathway databases [25]. A number of comprehensive maps of molecular interactions in the form of reaction networks representing various relatively large parts of molecular mechanisms, often disease-associated, were created following this way [11-19].

However, large maps of molecular interactions are difficult to exploit, maintain and improve without proper software support for navigation, querying and providing feedback on their content. There are a number of tools recently developed that take care of some of these aspects [26] (Table 1). Such tools as CellPublisher [27], Pathway Projector [28], GenMAPP [29] and PathVisio [30] focus on navigating within the maps, in particular, exploiting the geographical metaphor and using the Google Maps engine. SBGN-ED [31] supports all SBGN diagram types. WikiPathways [8] and Payao [32] focus on the web-based service for network annotation and curation. Similarly, PathBuilder is an example of a web-based pathway resource including an annotation tool [33]. The BioUML platform supports SBGN, SBML and enables the maps to connect to other databases as well as collective drawing of maps, similar to the principles of Google Docs. Nevertheless, from our practical experience of map creation, maintenance and curation, we identified the lack of a tool that would offer simultaneously a) easy web-based navigation through the content of comprehensive maps of molecular interactions created according to the systems biology standards; b) visualization of the map at different scales in a readable form; c) possibility to collect the feedback of a user about the map's content in an interactive manner. To fill this unoccupied but highly-demanded niche, we have developed NaviCell which uses the Google Maps engine for navigation into the map, semantic-zooming principles for exploring the map at various scales and the standard WordPress blogging system for collecting comments on the maps, providing a discussion forum for the community around the map's content. The combination of these three features makes NaviCell unique and a useful tool for enabling user-friendly navigation, curation and maintenance of (large) maps of molecular interactions. NaviCell is publicly available at http://navicell.curie.fr.

## Implementations

### NaviCell architecture and installation

NaviCell is a bioinformatics environment which allows the conversion of a large CellDesigner xml file into a set of images and html pages, containing Google Maps javascript code (Figure 1). These pages can be placed onto a web-server or used locally through all major flavors of Internet browsers. The procedure of creating map representations in the form of NaviCell pages is straightforward, and, in the simplest case of a browse-only representation, takes only a few clicks and several minutes. Creation of the blog is also straightforward but requires installation of the WordPress server and automatic generation of topics (posts) in the blog (Figure 1). NaviCell is supposed to be used by users with two roles: a) *a map manager* who creates, updates and annotates the map; and b) *a map user* who navigates the map through the web-interface, and comments its content through the blog.

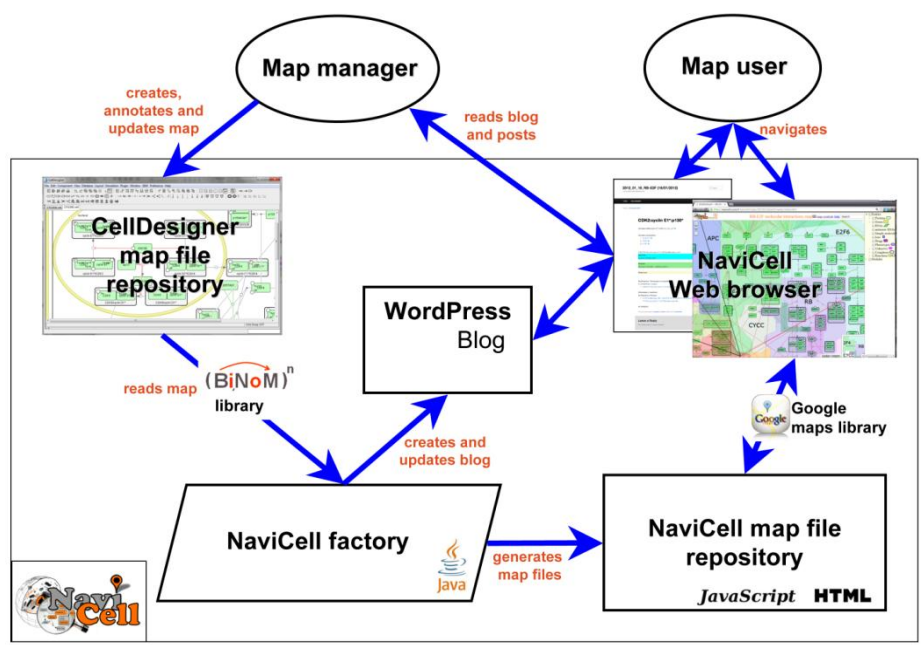

Figure 1. General architecture of NaviCell environment.

**NaviCell ontology**

Since NaviCell uses the CellDesigner files for creating their web-based representations, it reuses the ontology of biological entities implemented in CellDesigner SBML extension. NaviCell distinguishes Proteins, Genes, RNAs, antisense RNAs as distinct biological entities, existing with different modifications (eg., different phosphorylated forms of the same protein) and in different cell compartments. Moreover, the same modification of an entity can be represented on the map at several places by multiple aliases. NaviCell creates and displays in the selection panel an explicit list of all map elements and groups them per type of biological entities as described above (see Figure 2).

Naming the map elements in NaviCell is adopted from the BiNoM Cytoscape plugin [34]. More precisely, entity names are combined with other features such as modifications, compartment names and complex components. The different features are indicated by special characters, such as "@" for the compartments, "|" for modifications and ":" to delimit the different components of a complex. For example,

the name "Cdc25|Pho@cytoplasm" represents the protein Cdc25 in a phosphorylated state, located in the cytoplasm, while the name "Cdc13:Cdc2|Thr167_pho@cytoplasm" indicates a protein complex located in the cytosplasm, composed of the protein Cdc13 and the protein Cdc2 which is phosphorylated at position 167 on a threonine residue.

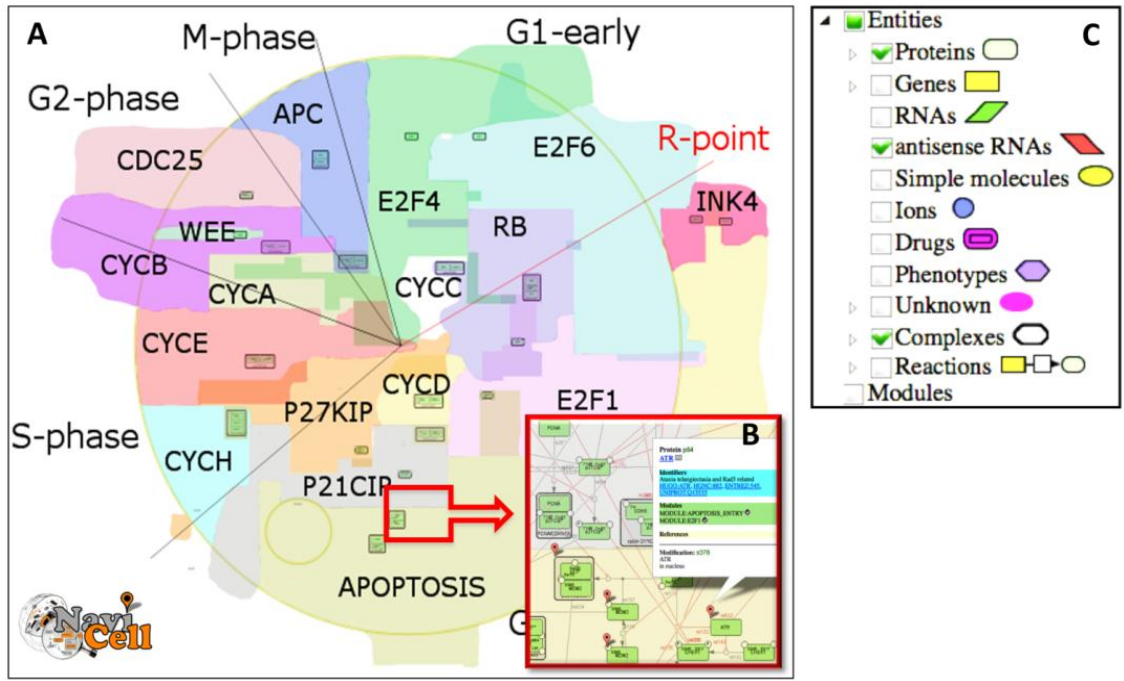

**Figure 2. NaviCell layout.** (A) Map panel with top-level view of the modular RB-E2F map, (B) Fragment of the detailed view of the map with pop-up bubble and markers, (C) Selection panel with list of the map entities grouped per type.

**Semantic zooming**

Semantic zooming is a visualization principle which is widely used in modern geographical maps and consists in providing readable image of the observed part of the map at each zoom level [35]. In semantic zooming, details of representation vanish progressively when zooming out, and individual objects or group of objects can be replaced by representations more suitable and informative for the current scale of visualization (see Figure 3 for an example). By contrast, simple mechanistic compression of pixels that is used currently in most biological network viewers makes network images uninformative. Because of this, the idea of semantic zooming in application for visualization of biological networks attracted recently some attention [36, 37]. In NaviCell, the map manager can provide as many semantic zooming views of the network as needed, and creation of the corresponding images remains in the manager's hands. The only requirements that should be met come from Google Maps: a) The size of the each semantic zooming view should be half-size of the previous view; b) Relative object position should not change in different zooming views; c) It is preferable that each image size (in pixels) of the semantic zooming could be divided by 256.

At the most-detailed zoom level, the image contains all the details, while at least-detailed zoom, the image provides a general top-level view of the map, hiding details and simplifying representations of individual objects. Therefore, semantic zooming consists in gradual hiding and transforming the details of information to give a meaningful abstract representation when zooming out from the detailed towards the top level view. These ideas are demonstrated in the Results section using the RB/E2F pathway map example.

In the simplest case, a user might provide only the most detailed and the least detailed (top-level) view of the map. NaviCell is able to generate the missing images automatically, by reducing the numbers of pixels of the images (not semantic zoom views).

**Preparing biological network maps for NaviCell**

*General requirements*

There are three necessary elements for generating the NaviCell representation of a comprehensive map of molecular interactions: a) a map file in CellDesigner xml format (the master map); b) a set of semantic zooming views of the map (in PNG format). At least two levels must be present: the most detailed and the top-level images; c) a simple configuration file, specifying several options for generating NaviCell files.

*Preparation of map modules*

In addition, a user can split the map (master map) into sub-maps called modules, typically defined on functional or structural basis, though any other criteria might be used. An unlimited number of separate simplified map representations that can contain subsets of the master map objects can accompany the master map in NaviCell. Each module can be represented with its own layout in the most clear and readable form. NaviCell allows accessing and shuttling between the map's modules. This option is of the utmost convenience for facilitating the map exploration as it is demonstrated in the Results section on the example of RB/E2F map.

*NaviCell annotation format*

NaviCell is capable to process a structured form of annotations of biological entities, if the CellDesigner xml file contains annotations in appropriate format. According to the NaviCell annotation format, annotations are structured in sections, each grouping specific information (in our examples, we used sections "Identifiers", containing the standard entity's IDs, "Modules" for specifying to which module a particular entity belongs, and "References" for providing links to publication records of the entity and free comments). Different sections are further highlighted by different colors in the NaviCell interface, making them easier to distinguish in the pop-up bubbles and in the blog posts. In addition, NaviCell converts several types of tags into hyperlinks such that a user could immediately access external databases such as HUGO, UniProt, Pubmed, etc.

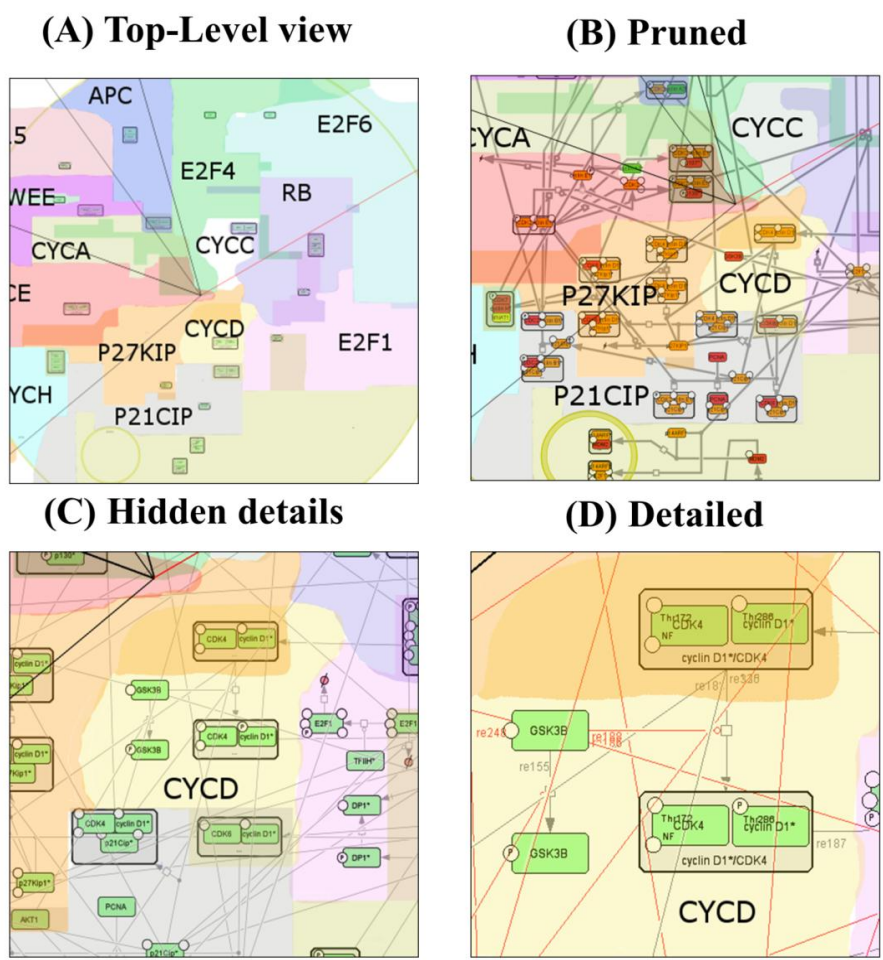

**Figure 3. NaviCell semantic zooming.** The same area of the map is visualized at 4 different zoom levels; each image is twice smaller than in the preceding zoom level. (A) In the top-level view, boundaries of map modules are visualized. (B) In the pruned view, only the most important molecular cascades are visualized. (C) In the hidden details view, unreadable tiny details (such as residue names) are hidden. (D) In the detailed view, entity names, modifications and reaction IDs are visible.

If the entities of the map are not annotated or annotated using a format differing from NaviCell's, the pop-up bubble and annotations in the blog are generated with raw annotations, without the sections that are described above.

*NaviCell map generation*

When the necessary files are prepared, the NaviCell map representation is generated through a menu "BiNoM/BiNoM I/O/Produce NaviCell maps files…" of the BiNoM Cytoscape plugin [34]. This is done through a simple dialog asking to indicate the location of the configuration file. There are two modes of generating NaviCell files.

The simple mode produces a local set of files with annotations as html files. In this case, no commenting on the map's content is possible. The complete mode requires pre-installed WordPress blogging system, creating and configuring a new blog devoted to the map, and specifying credentials for a user of WordPress with administration user rights, in order to automatically generate new posts in the blog. The source xml file of the map might be or not presented to users: this will depend on the policy of the map manager.

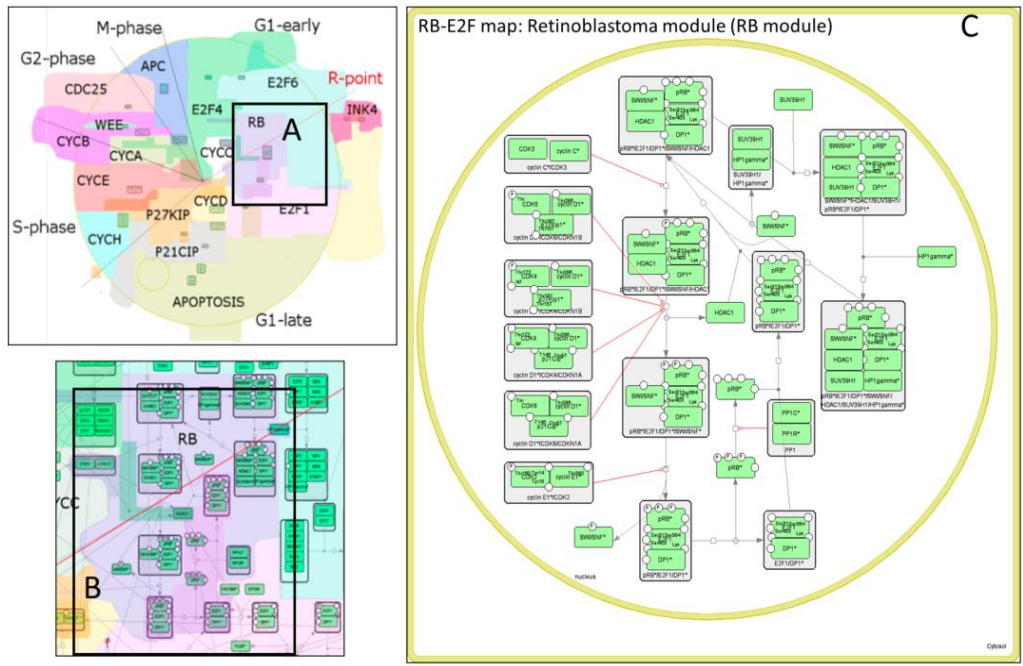

**Figure 4. Module maps.** (A) RB module on the Top-level view zoom, (B) RB module on the hidden-details view zoom, (C) RB module shown as a separate map.

*Collecting user's feedback and expertise*

A unique feature of NaviCell is the possibility to create and generate a web-blog for providing a discussion forum around an already existing map. The users browsing the map can view the blog entries without registering, or leave their comments after a simple registration procedure.

NaviCell implements a mechanism of updating the blog entries when the map is modified. If an entity in the map is removed, then the corresponding blog entry is moved to archive and becomes invisible. If the annotation of an entity in the map is changed, then the corresponding blog entry is updated, and the previous version is moved into the archive together with all previous users' comments. The blog archive is accessible to users, by this way providing traceability of all discussions and changes. This mechanism is implemented to avoid confusion in discussion if an entity's annotation has been significantly modified.

**Figure 5. Annotation post in the blog for the complex CDK2:cyclin A2*:p27Kip1***

**NaviCell documentation**

The NaviCell is accompanied by two detailed guides downloadable from the NaviCell web site. There is the 'Guide for map manager and system administrator' describing in details NaviCell installation procedure, recommendations for map construction and structuring entity annotations for the most efficient use of NaviCell, instructions for semantic zooming levels creation, recommendations for preparing map modules and using the map in NaviCell (http://navicell.curie.fr/pages/install.html).

For NaviCell users that are interested to explore and comment the existing maps without installing NaviCell and uploading their map to NaviCell, the explanations of NaviCell layout and instructions for efficient navigation and commenting maps in NaviCell can be found in the 'Guide for user and map curator' (http://navicell.curie.fr/pages/guide.html).

In addition, a video tutorial is available which provides a fast introduction to NaviCell features and demonstration of maps navigation and commenting in NaviCell (http://navicell.curie.fr/pages/tutorial.html).

# Results

**Example of RB/E2F map in NaviCell**

To illustrate NaviCell's capabilities, we use the map of RB/E2F pathway we created earlier [17]. We redesigned the global map layout which now reflects the cell cycle organization (Figure 2). We made four levels of semantic zooming (Figure 3) as described below in more details. The RB/E2F map contains 16 modules as they were described before [17]. Each module is also represented as a separate map with a simple and readable layout. The shuttling between the master map and the module maps is provided through internal links in NaviCell (Figure 4). The map is connected to a web-blog with pre-generated posts corresponding to each map's entity or module (Figure 5). Each such post provides full entity annotation specifying in what forms it exists, in what reactions it participates and which role it plays in reactions (reactant, product, catalyzer, etc.). The post can be commented by the map's users. A user can submit comments on the annotation posts in the form of hypertext enriched with images and hyperlinks. This blog is a system for knowledge exchange and active discussion between specialists in the corresponding domain and NaviCell map managers. The RB/E2F map is used for the video-tutorial of NaviCell's functions available at the web-site.

Together with the RB/E2F map, NaviCell representations are provided for the map of Notch and P53 pathway crosstalk that was used in one of our projects, as well as for number of large CellDesigner's maps that were published until so far in the literature [11, 13, 16, 18, 19]. The whole collection of maps is accessible from the NaviCell's web site http://navicell.curie.fr/pages/maps.html .

**Navigating a comprehensive map in NaviCell**

Navigation through the map of molecular interactions in NaviCell is assured by the standard and user-friendly engine of Google Maps, allowing scrolling, zooming, dropping down markers and showing pop-up bubbles (Figure 2). Using Google Maps makes it easy to get started with NaviCell, as it is an intuitive and widely used interface. The content of the map is detailed in the right-hand selection panel, which contains a list of entities and map objects grouped by their types. The selection panel allows identifying and selecting the desired map element and dropping down a clickable marker. Also a user can identify a molecular entity by using the full-text search function, which queries molecular entities by any substring in their description (by name, by synonym if provided, by word in the annotation). The markers do not disappear when zooming in or out, indicating the positioning of the selected map elements at all zoom levels. Clicking on a marker opens a pop-up bubble that contains a short description of the selected entity, hyperlinks to external databases; internal hyperlinks to related map objects and a link to the corresponding annotation post in the blog where the user's comments can be posted (see Figure 5 and below for the blog explanation).

**Semantic zooming views of the RB/E2F map in NaviCell**

The RB/E2F map can be navigated similarly to geographical maps through Google Maps, by several levels of map views (Figure 3). The navigation of the map starts from the top-level view (Figure 3A) where one can identify 16 map modules with their names (analogous to countries or regions), together with the positioning of the most important molecular entities inside them (analogue of country capitals). In addition, the cell cycle phases are indicated.

In the next level, a more detailed level of zooming called "Pruned" in Figure 3B, the major or canonical cell signaling pathways are visualized. These pathways were specified by intersecting the content of the RB/E2F map with several pathway databases and selecting those entities and reactions that are "canonically" represented in those databases (see the NaviCell guide for semantic zoom levels generation explanation).

The third zoom level is a "Hidden details" level shows the RB/E2F map with all molecular players and reactions. The tiny details such as the names of post-translational modification residues, complex names, and reaction ids are hidden at this level. In addition, the relative font sizes are increased for better readability (Figure 3C).

The fourth semantic zooming level is the most detailed view where all map elements are present (Figure 3D).

The module background coloring appears as a context layer in the background of all levels of zooming, which allows having an idea which modules of the map are currently visualized, at all zoom levels [38].

**RB/E2F map modules representation in NaviCell**

Each molecular entity of the RB/E2F map can be found in one or more of the 16 modules. The participation of an entity in various modules is indicated in the protein's annotation bubble by a link which leads to a separate module map(s). In addition, each module map can be accessed from the selection panel. The module map represents the corresponding sub-map with a simplified layout (Figure 4); the right hand panel contains the list of only those entities that are associated with this module. Thus, the master map is connected to a collection of maps of modules, and a user can easily shuttle between them. An example of this is shown in Figure 4: the RB module map is shown in Figure 4C, and it helps to understand the whole life cycle of the RB protein (how it gets phosphorylated at different residues, in which complexes it participates, what are the regulators of the main transitions).

**Blogging in NaviCell: commenting the RB/E2F map**

NaviCell uses the WordPress (http://wordpress.org) web-based blog system to collect feedback from the map users. The blog contains pre-generated posts for each entity of the map (genes, proteins, complexes, reactions etc.). Each post is composed of a detailed entity annotation (HUGO names, references, etc.), links to other entities in the network (internal hyperlinks) and links to databases including PubMed, NCBI etc. (external hyperlinks).

An example of a pre-generated post is shown in Figure 5 for a molecular complex composed of CDK6, Cyclin D1 and p27Kip1 proteins. The post contains annotation of the complex, list of complex forms (modifications) and reactions in which the complex participates as reactant or product, or catalyzer. Note that each globe icon in the post leads to the map, and allows selecting corresponding objects on it. For example, clicking at the globe icons in the line describing the reaction DP2*:E2F4:p107*@nucleus → DP2*:E2F4:p107*|pho@nucleus, the markers will show either DP2*:E2F4:p107*@nucleus species or DP2*:E2F4:p107*|pho@nucleus species or the reaction itself on the map. Parallel use of the map and the blog facilitates exploring and understanding the map.

The blog system provides a feedback mechanism between the map users and map managers. Updating the map is foreseen in the following scenario. The manager of the map regularly collects the users' comments and updates the map accordingly in a series of releases. In turn, NaviCell can automatically update the blog and archive older versions of posts including users' comments, thus providing traceability of all changes on the map and simplifying map maintenance (See Figure 1).

## Discussion

NaviCell is an environment for visualization and simplified usage of large-scale maps of molecular interactions created in CellDesigner. NaviCell allows showing the content of the map in a convenient way, at several scales of complexity or abstraction; it provides an opportunity to comment on its content, facilitating maintenance of the maps. NaviCell is not implemented to cover the functionality of all existing network visualization tools: however, NaviCell combines several essential features together, and therefore fits to a quite demanded niche in the map maintenance and support process.

The use of the Google Map interface makes it straightforward for the user to get started with NaviCell, as this interface is intuitive and already familiar to most users.

The development and application of semantic zooming principles is a unique feature of NaviCell that allows step-wise exploration of the map and helps to grasp the content of very complex maps of molecular interactions at several levels of complexity from the global map structure, through major, canonical pathways up to the most detailed level.

The concept of blog which we deliberately used for collecting the users' feedback in NaviCell competes with another paradigm of community-based map construction represented by WikiPathways [8]. In this model map editing is open to everybody, whereas in our model a map manager is in charge of validating propositions of modifications. In many cases, comprehensive maps are a one person project requiring thoughtful design of the map's layout and resolving contradicting interpretations of biochemical experiments and points of view. NaviCell, unlike WikiPathways, is not designed for collective *ab initio* construction of the maps but, instead, allows visible and open discussion forum around an already existing map. Later the map can be modified and updated accordingly by the map manager who takes responsibility and interprets the users' comments, not allowing uncontrolled map changes. We believe that both paradigms (blog *vs* wiki) are of interest in the systems biology field, and can be combined in the future.

We believe that NaviCell will reinforce the interest to assemble large-scale maps of molecular interactions and present them to the community for constructive critics. We hope that in such a way more consensual formalizations of large maps of molecular mechanisms will be achieved.

We now plan to extend NaviCell with an analytical toolbox implementing a set of methods for visualizing high-throughput data (expression measurements, protein activities, mutation profiles, etc.) on top of the molecular maps, and with tools for analyzing the map's structure in the spirit of Google Maps (for example, route finding, suggesting several alternative routes, finding the ways to cut the routes).

## Conclusions

NaviCell is a web-based, user-friendly and interactive environment, which can be easily used by molecular biologists. NaviCell functionality has been already tested in several concrete projects for navigation and curation of large maps of molecular interactions.

In the future, we plan to convert NaviCell into an environment under which maps will be visualized, discussed, updated and analyzed using a toolbox that is currently in development.

## Availability and requirements

Project name: NaviCell

Project homepage: http://navicell.curie.fr

Operating system(s): Platform independent

Programming language: Java, HTML, JavaScript

License: GNU GPL

Any restrictions to use by non-academics: license is not needed

# Competing interests

The authors declare that they have no competing interests.

# Authors' contrubution

The project was conceived by AZ, IK and EB. AZ developed the algorithms and coordinated setting up the environment. SP developed the software code and participated in setting up the environment. IK coordinated the project. DC, LC, IK and AZ prepared maps in the NaviCell format. DC and IK designed and implemented the web site. IK, DC, SP and AZ. prepared documentation and guidelines. IK and AZ have written the manuscript. All authors read and approved the final manuscript.

# Acknowledgments

We thank Camille Barette for help in setting up the informatics recourses and Hien-Anh Nuigen for help with NaviCell guides design. This work is supported by the APO-SYS EU FP7 project under grant n° HEALTH-F4-2007-200767, by the grant INCA LABEL Cancéropole Ile-de-France 2011-1-LABEL-1, by the grant INVADE from ITMO Cancer (Call Systems Biology 2012) and by the grant "Projet Incitatif et Collaboratif Computational Systems Biology Approach for Cancer » from Institut Curie. IK, DC, LC, EB and AZ are members of the team "Computational Systems Biology of Cancer", Equipe labellisée par la Ligue Nationale Contre le Cancer.

| Feature | Map browsers used in pathway databases ||||| Standalone tools |||||||||
| --- | --- | --- | --- | --- | --- | --- | --- | --- | --- | --- | --- | --- | --- | --- | --- |
| | KEGG | Reactome | Spike | Wiki Pathways | Panther | NaviCell | Cell Publisher | Pathways Projector | Cell Designer | Payoa | SBGN-ED | GenMAPP | Pathviseo | BioUML | Cytoscape +plugins |
| SBGN representation (or close to SBGN) | | | | | × | × | × | | × | × | × | | | × | × |
| Scalable to large maps | N/A | N/A | N/A | N/A | N/A | × | × | × | × | | × | N/A | N/A | × | × |
| Modular map representation | | × | | | | × | | | × | | × | | | × | × |
| Structured annotation of entities | × | × | | × | × | × | × | × | | | | × | × | × | × |
| Map editing | | | | × | | | | | × | | × | | × | × | |
| Dynamic layout | | | | | | | | | × | | × | | | × | × |
| Analytical tools | ×* | ×* | | | ×* | | | ×* | ×* | | ×* | ×* | ×* | ×* | ×* |
| Web interface | × | × | | × | × | × | × | × | | | | | | × | |
| GoogleMaps-like interface | | | | × | | × | × | × | | | | | | | |
| Semantic zooming | | × | | | | × | | | | | | | | | |
| Full text search | × | × | | × | × | × | | × | | | | | | × | |
| Links to databases | × | × | | × | × | × | × | × | × | | × | × | × | × | |
| Associated discussion forum | | | | × | | × | | | | × | | | | | |
| Archiving discussions | | | | × | | × | | | | × | | | | | |
| Browser | Any type | Any type | Any type | Any type | Any type | Firefox Safari Chrome | Firefox Safari Chrome, IE | Firefox Safari | N/A | N/A | N/A | N/A | N/A | Any type | N/A |
| Reference | 5 | 4 | 7 | 8 | 6 | | 27 | 28 | 22 | 32 | 31 | 29 | 30 | | 34 |

**Table 1  Comparison of NaviCell features with existing tools**

*****Analytical tools:**

**Pathway projector-**Data mapping, Path search, Sequence-based search (BLAST), Pathway prediction**; KEGG-**Data mapping, Pathways enrichment analysis**; Reactome-**Data mapping, Pathways enrichment analysis, Path search; **CellDesigner-**Simulation; **SBGN-ED-**Data mapping, Network analysis, Simulation; **Panther-**Data mapping; **PathVisio-**Data mapping, Pathways enrichment analysis**; GenMAPP-**Data mapping, Pathways enrichment analysis**; BioUML-**Simulation and analysis; **Cytoscape**-Various tools available through plugins.